\documentclass{elsart}
\usepackage{graphicx}
\renewcommand{\vec}[1]{\mbox{\boldmath ${#1}$}}
\begin{document}
\begin{frontmatter}
\title{Can the Landau-Lifshitz equation explain the  spin-wave instability in
 ferromagnetic thin films for parallel pumping?}
\author{Kazue Kudo\corauthref{cor}}, 
\corauth[cor]{Corresponding author. Tel.: +81 6 6605 2768; 
 fax: +81 6 6605 2768.}
\ead{kudo@a-phys.eng.osaka-cu.ac.jp}
\author{Katsuhiro Nakamura}
\address{Department of Applied Physics, Osaka City University,
              Osaka 558-8585, Japan }
\begin{abstract}
Spin-wave instability is studied analytically in the case of parallel
 pumping for thin films under external field perpendicular to the film
 plane. It is examined whether the instability threshold derived from 
 only the Landau-Lifshitz (LL) equation can explain 
 experimental instability threshold without using the
 microscopically-derived spin-wave line 
 width $\Delta H_k$, which is conventionally used.
 It is revealed that the butterfly curve cannot be explained from only
 the LL equation  at least in an
 analytical way.
By contrast, for the case of 
 perpendicular pumping, the Suhl instability was well
 explained from the LL equation.
 The difference between the two cases comes from the nonlinear terms
 describing the relaxation of spin waves. It is suggested how the
 nonlinear terms in the LL equation should be related to  $\Delta H_k$
 for parallel pumping.
\end{abstract}
\end{frontmatter}

\section{Introduction}

Spin-wave instability is a nonlinear phenomena which is dominated by a
balance between excitation and relaxation of spin waves. In early 1950s,
anomalous phenomena incompatible with the conventional theory were
observed in perpendicular-pumping ferromagnetic resonance (FMR)
experiments~\cite{Damon,Bloem}.  The phenomena was successfully explained by
Suhl's theory~\cite{Suhl}. In the theory, the coupling between the
uniform mode ($\vec{k}=0$) and other spin-wave
modes ($\vec{k}\ne 0$) plays a crucial
role. In perpendicular pumping, the microwave magnetic field excites
the uniform mode. Energy is fed from a $\vec{k}=0$ mode or a pair of 
 $\vec{k}=0$ modes to a pair of spin waves $\vec{k}$ and $-\vec{k}$
 ($\vec{k}\ne 0$).
The pair of spin waves which is most strongly coupled to the uniform
mode grows exponentially when the intensity of the pumping field
exceeds a certain threshold.
Such processes can be well described by using only the Landau-Lifshitz (LL)
equation governing the magnetization dynamics. 

On the other hand, the spin-wave instability for parallel pumping has
been investigated in order to study relaxation phenomena of spin waves
with definite wave vectors since the first
experimental studies in
1960s~\cite{Schlomann,Morgen}. 
In the parallel pumping experiments, the microwave magnetic field is applied
parallel to the equilibrium magnetization. Therefore the longitudinal
component of a standing spin-wave, which consists of a pair of
traveling spin-waves $\vec{k}$ and $-\vec{k}$, interacts with the
microwave magnetic field. When the power of the pumping field exceeds a
certain threshold, the spin waves $\vec{k}$ and $-\vec{k}$ grow
exponentially. The threshold depends on the balance between excitation and
relaxation of spin waves. The relaxation of spin waves can be described
by the spin-wave line width $\Delta H_k$ which is derived from
microscopic magnon-magnon scattering theories. 
To explain the experimentally-observed  spin-wave
instability threshold, i.e. so-called butterfly curve, $\Delta H_k$ is
used as a factor of damping in most of conventional theories.
For example, Patton {\it et al.} proposed some trial $\Delta H_k$
functions and successfully explained butterfly curves of the instability
threshold~\cite{Patton}.
In contrast to the Suhl instability in the perpendicular pumping,
however,
there has been no attempt to use only the LL equation 
for explaining the experimental butterfly curve. 

In the 1980s, there occurred a renaissance on studies of 
FMR~\cite{Aguiar,Bryant,Carrol}. High-resolution experiments were
carried out for spin-wave nonlinear dynamics in yttrium iron garnet (YIG)
films and spheres in parallel and perpendicular pumping methods.
For certain high power beyond the Suhl instability threshold, the
interaction between excited spin-wave modes lead to various dynamical
phenomena such as chaos and turbulence. Since then the studies on spin
wave instabilities have acquired a renewed interest up to
now~\cite{Becker,Laulicht,LaulPTP}. Despite these pioneering experiments, the
corresponding theoretical investigations for parallel pumping leave
several crucial questions. In particular, no dynamical equation
leading to auto-oscillations and chaos~\cite{Lvov} and pattern
formations~\cite{Elmer} can touch on even 
a qualitative feature of the experimental butterfly curve for the instability
threshold.  In the case of parallel pumping, therefore, it is
important to analyze the butterfly curve from a viewpoint of modern
nonlinear dynamics.

In this paper, we try to derive analytically 
the spin-wave instability threshold in
ferromagnetic thin films for parallel pumping without using the
spin-wave line width $\Delta H_k$ which comes from microscopic theories.
By using only the LL equation, besides the linear instability
analysis, the nonlinear instability is examined for a simple case. 
It will be shown that the instability threshold obtained in that way cannot
explain experimental butterfly curves. We will reveal why our nonlinear
instability analysis fails and suggest what terms arising from
 the LL equation should correspond to $\Delta H_k$.

\section{Equations of motion for the spin-wave amplitudes}

The LL equation is given by
\begin{equation}
\frac1{\gamma}\partial_t \vec{M} =\vec{M}\times\vec{H}_{\rm eff}
 -\frac{\lambda}{M_0}\vec{M}\times (\vec{M}\times\vec{H}_{\rm eff}).
 \label{eq:LL} 
\end{equation}
Here, $\vec{M}(\vec{r})$ is a magnetization field and $M_0$ is its value
in thermal equilibrium;  $\gamma$ is the gyromagnetic ratio ($\gamma <0$
for spins); $\lambda$ is a damping constant
parameter ($\lambda <0$ in this case);
$\vec{H}_{\rm eff}$ is the effective magnetic field:
\begin{equation}
\vec{H}_{\rm eff}=D\nabla^2\vec{M} +\vec{H}^{\rm d}+\vec{H}^{\rm a}
+\vec{H}_0 +\vec{h}\cos\omega t.
\label{eq:Heff}
\end{equation}
The first term comes from the exchange interaction; 
the second term is the demagnetization term;
the third term is
the anisotropy field, which is omitted below for convenience;
the fourth and fifth
terms are the external static and pumping fields, respectively. 
We assume that $\vec{H}_0 // \vec{h}$ and these fields are parallel
to the $z$ axis.  
The demagnetization field is given as the gradient of the magneto-static
potential $\phi$:
\begin{equation}
 \vec{H}^{\rm d}=-\vec{\nabla}\phi .
\end{equation}
The magneto-static potential obeys the Poisson equation:
\begin{equation}
 \nabla^2\phi = \left\{
\begin{array}{rl}
 4\pi\vec{\nabla}\cdot\vec{M}, & \mbox{inside the sample}; \\
 0, & \mbox{outside the sample}.
\end{array}
\right.
\label{eq:poi}
\end{equation}

Throughout the text, we call a set of Eqs.~(\ref{eq:LL})-(\ref{eq:poi})
as the LL equation.
Now we introduce the normalized magnetization $\vec{S}=\vec{M}/M_0$,
considering the length of $\vec{M}$ is invariant: $M_0=|\vec{M}|$.
It is convenient to project $\vec{S}$ stereographically onto a complex
variable $\psi (\vec{r},t)$~\cite{Laksh}:
\begin{equation}
 \psi=\frac{S_x+\mathrm{i}S_y}{1+S_z},
\end{equation}
where
\begin{equation}
 S_x=\frac{\psi+\psi^*}{1+\psi\psi^*}, \quad
 S_y=\frac{\mathrm{i}(\psi^* -\psi)}{1+\psi\psi^*}, \quad
 S_z=\frac{1-\psi\psi^*}{1+\psi\psi^*}.
\label{eq:Ss}
\end{equation}
In terms of $\psi$, Eq.~(\ref{eq:LL}) is rewritten as
\begin{eqnarray}
 \partial_t \psi -\mathrm{i}(1-\mathrm{i}\lambda )&\gamma&
 \left\{ D M_0 \left[ \nabla^2\psi
 -\frac{2\psi^*(\nabla\psi)^2}{1+\psi\psi^*} \right]
 -\frac12 (1-\psi^2)\partial_x\phi \right. \nonumber\\
 &-&\left. \frac{\rm i}2 (1+\psi^{*2})\partial_y\phi
 +(\partial_z\phi-H_0-h\cos\omega t)\psi
\right\} =0.
\label{eq:LLs}
\end{eqnarray}
Inside the sample, $\phi$ satisfies
\begin{eqnarray}
 \nabla^2\phi&=&\frac{4\pi M_0}{(1+\psi\psi^*)^2}\left\{
 (1-\psi^{*2})\partial_x\psi+(1-\psi^2)\partial_x\psi^* \right.\nonumber\\
 &+&\left. \mathrm{i}[(1+\psi^2)\partial_y\psi^*-(1+\psi^{*2})\partial_y\psi ]
 -2(\psi\partial_z\psi^* +\psi^*\partial_z\psi) \right\}.
\label{eq:pois}
\end{eqnarray}

The demagnetizing field is affected by boundary conditions.
We consider a film of thickness $d$ infinitely extended in the $x$-$y$
plane under the external filed applied parallel to $z$ axis.
We assume unpinned surface spins, which satisfy Neumann-like boundary
conditions, 
\begin{equation}
 \left. \partial_z \vec{S} \right|_{z=\pm d/2}=0,
\end{equation} 
namely,
\begin{equation}
 \left. \partial_z \psi\right|_{z=\pm d/2}=
 \left. \partial_z \psi^*\right|_{z=\pm d/2}=0.
\label{eq:BC}
\end{equation}
Introducing the dimensionless time and space
units~\cite{Elmer}, 
\begin{equation}
 t\to \frac{t}{4\pi |\gamma |M_0}, \quad \vec{r}\to\vec{r}d,
\end{equation}
we obtain the linearized equations of motion corresponding to
Eqs.~(\ref{eq:LLs}) and (\ref{eq:pois}):
\begin{equation}
\partial_t\psi +\mathrm{i}(1-\mathrm{i}\lambda)\left[
l^2\nabla^2\psi -\frac12(\partial_x +\mathrm{i}\partial_y)\Phi
+(\partial_z\Phi -\omega_H -\omega_h\cos\omega_{\rm p}t)\psi
\right] =0,
\label{eq:LLl}
\end{equation}
\begin{equation}
 \nabla^2\Phi =(\partial_x -\mathrm{i}\partial_y)\psi
 +(\partial_x +\mathrm{i}\partial_y)\psi^* \quad
 \mbox{when $-\frac12 < z < \frac12$},
\label{eq:poil}
\end{equation}
where
\begin{equation}
 l^2=\frac{D}{4\pi d^2}, \quad \Phi =\frac{\phi}{4\pi M_0d}, \quad
 \omega_H =\frac{H_0}{4\pi M_0}, \quad \omega_h =\frac{h}{4\pi M_0},
 \quad \omega_{\rm p}=\frac{\omega}{4\pi M_0|\gamma |}.
\end{equation}

First of all, we consider the undriven case (i.e. $\omega_h=0$) and
expand $\psi (\vec{r},t)$ and $\psi^* (\vec{r},t)$ so that they fulfills
the boundary conditions~(\ref{eq:BC}). 
For even modes, $k_z=2m\pi$ ($m$: integer),
\begin{eqnarray}
 \psi (\vec{r},t)&=&\sum_{\vec{\scriptstyle k}}
 a_{\vec{\scriptstyle k}}(t)
 \e^{\mathrm{i}(k_x x+k_y y)}
 \cos k_z z \nonumber\\
 \psi^* (\vec{r},t)&=&\sum_{\vec{\scriptstyle k}}
 a^*_{-\vec{\scriptstyle k}}(t)
 \e^{\mathrm{i}(k_x x+k_y y)}
 \cos k_z z.
\label{eq:expan_a}
\end{eqnarray}
For odd modes, $k_z=(2m+1)\pi$ ($m$: integer),
\begin{eqnarray}
 \psi (\vec{r},t)&=&\sum_{\vec{\scriptstyle k}}
 a_{\vec{\scriptstyle k}}(t)
 \e^{\mathrm{i}(k_x x+k_y y)}
 \sin k_z z \nonumber\\
 \psi^* (\vec{r},t)&=&-\sum_{\vec{\scriptstyle k}}
 a^*_{-\vec{\scriptstyle k}}(t)
 \e^{\mathrm{i}(k_x x+k_y y)}
 \sin k_z z.
\label{eq:expan_b}
\end{eqnarray}
Using the expansions~(\ref{eq:expan_a}) and (\ref{eq:expan_b}), we obtain
solutions of Eq.~(\ref{eq:poil}): 
\begin{eqnarray}
 \Phi (\vec{r})&=&-\mathrm{i}\sum_{\vec{\scriptstyle k}}
 \frac{\e^{\mathrm{i}(k_x x+k_y y)}}{k^2}
(k_{-}a_{\vec{\scriptstyle k}} +k_{+}a^*_{-\vec{\scriptstyle k}})
f_{\vec{\scriptstyle k}}\cos k_z z 
\quad \mbox{for even modes;} \nonumber\\
 \Phi (\vec{r})&=&-\mathrm{i}\sum_{\vec{\scriptstyle k}}
 \frac{\mathrm{e}^{\mathrm{i}(k_x x+k_y y)}}{k^2}
(k_{-}a_{\vec{\scriptstyle k}} -k_{+}a^*_{-\vec{\scriptstyle k}})
f_{\vec{\scriptstyle k}}\sin k_z z 
\quad \mbox{for odd modes.}
\label{eq:Phi}
\end{eqnarray}
Here $k_{+}=k_x+\mathrm{i}k_y$, $k_{-}=k_x-\mathrm{i}k_y$, 
$k_\perp =\sqrt{k_x^2+k_y^2}$ and 
\begin{eqnarray}
 f_{\vec{\scriptstyle k}}&=&1-\left(
1-\e^{-k_{\perp}} \right) \frac{k_{\perp}}{k^2}
\quad \mbox{when $k_z=0$;} \nonumber\\
 f_{\vec{\scriptstyle k}}&=&1-2\left(
1-\e^{-k_{\perp}} \right) \frac{k_{\perp}}{k^2}
\quad \mbox{when $k_z\ne 0$.} \nonumber\\
\end{eqnarray}
The detailed derivation of these solutions is shown in
Appendix~\ref{sec:app} and Ref.~\cite{Kudo}.
The derivatives $\partial_x\Phi$ and $\partial_y\Phi$ are calculated
from Eq.~(\ref{eq:Phi}). However, $\partial_z\Phi=-H^{\rm d}_z/4\pi M_0$ 
is approximated 
with the value of uniform magnetization, $\vec{k}=0$: 
$H^{\rm d}_z=-4\pi N_z M_z$, where $N_z$ is a demagnetizing factor.
In this case, $N_z=1$. Considering $M_z\simeq M_0$, we have 
$\partial_z\Phi=1$. 
Then Eq.~(\ref{eq:LLl}) for $\omega_h=0$ is rewritten as
\begin{equation}
\partial_t\psi +\mathrm{i}(1-\mathrm{i}\lambda)\left[
l^2\nabla^2\psi -\frac12(\partial_x+\mathrm{i}\partial_y)\Phi
+(1 -\omega_H )\psi \right] =0.
\label{eq:LLl2}
\end{equation}
From Eqs.~(\ref{eq:expan_a})-(\ref{eq:LLl2}), the linearized equation of
motion of $a_{\vec{\scriptstyle k}}$ is found to be
\begin{equation}
\partial_t a_{\vec{\scriptstyle k}}
 +\mathrm{i}(1-\mathrm{i}\lambda) A_{\vec{\scriptstyle k}}
   a_{\vec{\scriptstyle k}}
 +\mathrm{i}(1-\mathrm{i}\lambda) B_{\vec{\scriptstyle k}}
   a^*_{-\vec{\scriptstyle k}}=0,
\label{eq:al}
\end{equation}
where
\begin{eqnarray}
A_{\vec{\scriptstyle k}}&=&1 -l^2 k^2 -\omega_H-\frac12
f_{\vec{\scriptstyle k}}\sin^2\theta_{\vec{\scriptstyle k}},\nonumber\\
B_{\vec{\scriptstyle k}}&=&-\frac{(-1)^n}2 
\e^{2\mathrm{i}\varphi_{\vec{\scriptscriptstyle k}}}
f_{\vec{\scriptstyle k}}\sin^2\theta_{\vec{\scriptstyle k}},\nonumber\\
\sin\theta_{\vec{\scriptstyle k}}&=&\frac{k_\perp}{k}, \quad
\exp (\mathrm{i}\varphi_{\vec{\scriptstyle k}}) =
\cos\varphi_{\vec{\scriptstyle k}}+
\mathrm{i}\sin\varphi_{\vec{\scriptstyle k}}=
\frac{k_x}{k_\perp}+\mathrm{i}\frac{k_y}{k_\perp}.
\label{eq:AB}
\end{eqnarray}
Here, $n$ is an integer and $k_z=\pi n$.

Equation~(\ref{eq:al}) represents two coupled harmonic oscillators
$a_{\vec{\scriptstyle k}}$, $a^*_{-\vec{\scriptstyle k}}$ and
can be diagonalized by means of the Holstein-Primakoff
transformation:
\begin{eqnarray}
 a_{\vec{\scriptstyle k}}&=&
 \nu_{\vec{\scriptstyle k}}b_{\vec{\scriptstyle k}}
-\mu_{\vec{\scriptstyle k}}b^*_{-\vec{\scriptstyle k}}\nonumber\\
 a^*_{-\vec{\scriptstyle k}}&=&
 \nu_{\vec{\scriptstyle k}}b^*_{-\vec{\scriptstyle k}}
-\mu^*_{\vec{\scriptstyle k}}b_{\vec{\scriptstyle k}},
\label{eq:H-P}
\end{eqnarray}
where
\begin{eqnarray}
  \nu_{\vec{\scriptstyle k}}&=&
 \cosh\frac{\chi_{\vec{\scriptstyle k}}}{2},\quad
 \mu_{\vec{\scriptstyle k}}=
 \e^{\mathrm{i}\beta_{\vec{\scriptscriptstyle k}}}
 \sinh\frac{\chi_{\vec{\scriptstyle k}}}{2},\nonumber\\
 \cosh\chi_{\vec{\scriptstyle k}}&=&
 \frac{\left| A_{\vec{\scriptstyle k}}\right| }
{\left[ A_{\vec{\scriptstyle k}}^2-(1+\lambda^2)
\left| B_{\vec{\scriptstyle k}}\right|^2 \right]^{1/2}}.
\end{eqnarray}
After the transformation (\ref{eq:H-P}), Eq.~(\ref{eq:al}) becomes
\begin{equation}
  \partial_t b_{\vec{\scriptstyle k}} -\mathrm{i}
(\omega_{\vec{\scriptstyle k}}+\mathrm{i}\eta_{\vec{\scriptstyle k}})
 b_{\vec{\scriptstyle k}}=0,
\end{equation}
where
\begin{eqnarray}
 \omega_{\vec{\scriptstyle k}}^2 &=&A_{\vec{\scriptstyle k}}^2
 -(1+\lambda^2) \left| B_{\vec{\scriptstyle k}}\right|^2,
 \label{eq:disp} \\
 \eta_{\vec{\scriptstyle k}}&=&\lambda A_{\vec{\scriptstyle k}}
\label{eq:damp}
\end{eqnarray}
Equations~(\ref{eq:disp}) and (\ref{eq:damp}) express the dispersion
relation and a damping rate for the spin wave, respectively.

\section{Linear instability}

We will first analyze the linear instability for the spin wave under the
driving field.
When the pumping field $\vec{h}\cos\omega t$ is applied, the equation of
motion of $a_{\vec{\scriptstyle k}}$ 
corresponding to Eq.~(\ref{eq:al}) becomes 
\begin{equation}
 \partial_t a_{\vec{\scriptstyle k}}
 +\mathrm{i}(1-\mathrm{i}\lambda) A_{\vec{\scriptstyle k}}
   a_{\vec{\scriptstyle k}}
 +\mathrm{i}(1-\mathrm{i}\lambda) B_{\vec{\scriptstyle k}}
   a^*_{-\vec{\scriptstyle k}}
 -\mathrm{i}(1-\mathrm{i}\lambda) \omega_h\cos(\omega_{\rm p}t)
  a_{\vec{\scriptstyle k}}=0.
\label{eq:ap}
\end{equation}
After the the transformation (\ref{eq:H-P}),
we substitute the following equations into Eq.~(\ref{eq:ap}),
\begin{eqnarray}
 b_{\vec{\scriptstyle k}}(t)&=&b^o_{\vec{\scriptstyle k}}(t)
\exp [\mathrm{i}(\omega_{\rm p}/2)t-\eta_{\vec{\scriptstyle k}}t],
\nonumber\\
 b^*_{-\vec{\scriptstyle k}}(t)&=&b^{o*}_{-\vec{\scriptstyle k}}(t)
\exp [-\mathrm{i}(\omega_{\rm p}/2)t-\eta_{\vec{\scriptstyle k}}t],
\label{eq:bo}
\end{eqnarray}
since the resonance occurs at
$\omega_{\vec{\scriptstyle k}}=\omega_{\rm p}/2$.
When only terms contributing to the resonance are left,
the variable $b^o_{\vec{\scriptstyle k}}$ satisfies
\begin{equation}
 \partial_t^2 b^o_{\vec{\scriptstyle k}} +
 \left[ \left( \omega_{\vec{\scriptstyle k}}
 -\frac{\omega_{\rm p}}{2}\right)^2
 - \left| \rho_{\vec{\scriptstyle k}} \right|^2 \right]
 b^o_{\vec{\scriptstyle k}}=0,
\label{eq:bp}
\end{equation}
where
\begin{equation}
 \left|  \rho_{\vec{\scriptstyle k}} \right| =
 \omega_h\sqrt{1+\lambda^2}
 \frac{\left| B_{\vec{\scriptstyle k}}\right| }
 {2\omega_{\vec{\scriptstyle k}}}.
\label{eq:rho}
\end{equation}
Therefore, the exponentially increasing solution for
$b_{\vec{\scriptstyle k}}$ is
\begin{equation}
 b_{\vec{\scriptstyle k}}\propto \exp \left[
\left( \left| \rho_{\vec{\scriptstyle k}} \right|
-\eta_{\vec{\scriptstyle k}} \right)t +\mathrm{i}(\omega_{\rm p}/2)
t \right],
\label{eq:bprop}
\end{equation}
where
\begin{equation}
 \left| \rho_{\vec{\scriptstyle k}} \right| > \eta_{\vec{\scriptstyle k}}.
\end{equation}
From Eqs.~(\ref{eq:rho}) and (\ref{eq:bprop}), the instability threshold 
$\omega_h^{\rm crit}$ is given as
\begin{equation}
  \omega_h^{\rm crit}=\frac{\omega_{\rm p}}{\sqrt{1+\lambda^2}}
 \min_{\vec{\scriptstyle k}} \left\{
 \frac{\eta_{\vec{\scriptstyle k}}}
{\left| B_{\vec{\scriptstyle k}} \right| } \right\}.
\label{eq:dam_dH}
\end{equation}
Let us rewrite Eq.~(\ref{eq:dam_dH}) with use of Eqs.~(\ref{eq:disp}) and
(\ref{eq:damp}): 
\begin{equation}
  \omega_h^{\rm crit}=\frac{\omega_{\rm p}\lambda}{\sqrt{1+\lambda^2}}
 \min_{\vec{\scriptstyle k}} \left[
 \frac{\omega_{\rm p}^2}
{4\left| B_{\vec{\scriptstyle k}} \right|^2 } +1+\lambda^2
\right]^{1/2}.
\label{eq:dam_dH2}
\end{equation}

By using Eqs.~(\ref{eq:disp}) and (\ref{eq:dam_dH2}), we might see a
theoretical butterfly curve, which is an instability threshold curve
plotted against the static field. However, the butterfly curve calculated
in that way is found to be totally different from experimental ones. 
Typical butterfly curves observed in experiments have a cusp at a
certain static field: 
as the static field increasing, $\omega_h^{\rm crit}$ first decreases
up to the cusp point and then increases above that point.
Here let us focus on the case for static fields below the cusp point.
For those static field, the minimum threshold mode corresponds
to a spin wave propagating with $\theta_{\vec{\scriptstyle k}}=\pi /2$,
i.e. parallel to the film plane.
As the static field increases, the wave vector $\vec{k}$ of the threshold
modes decreases, and $k \simeq 0$ at the cusp.
On the contrary, from Eq.~(\ref{eq:AB}), one notices that  
$\omega_h^{\rm crit}$ in  Eq.~(\ref{eq:dam_dH2}) 
depends only on $f_{\vec{\scriptstyle k}}$ when
$\theta_{\vec{\scriptstyle k}}$ is fixed and that it grows as the static
field increases up to the cusp point. 
In fact, for $\theta_{\vec{\scriptstyle k}}=\pi /2$, 
\begin{equation}
 f_{\vec{\scriptstyle k}}=1-(1-\mathrm{e}^{-k})\frac{1}{k}
\end{equation}
because $k_z=0$ and $k_{\perp}=k$. As $k\to 0$, 
$f_{\vec{\scriptstyle k}}\to 0$ and then $\omega_h^{\rm crit}\to\infty$.
Such a divergence of $\omega_h^{\rm crit}\to\infty$ has not been 
observed in experiments. Namely, the theoretical threshold
$\omega_h^{\rm crit}$ in  Eq.~(\ref{eq:dam_dH2}) 
fails to explain the experimental butterfly curve.

One may say that the linear analysis is not sufficient to discuss
the spin-wave instability threshold. 
In the next section, several nonlinear terms are included in equation of
motion, and a nonlinear instability is analyzed with expectation to
overcome the above problem.

\section{Nonlinear instability}

Let us rewrite Eqs.~(\ref{eq:LLs}) and (\ref{eq:pois})
up to third order of $\psi$. Then we have
\begin{eqnarray}
 \partial_t\psi +\mathrm{i}(1-\mathrm{i}\lambda )\left\{ l^2
\left[ \nabla^2\psi -2\psi^*(\nabla\psi )^2 \right] \right.
&-&{\textstyle \frac12}(\partial_x+\mathrm{i}\partial_y)\Phi
+{\textstyle \frac12}\psi^2(\partial_x-\mathrm{i}\partial_y)\Phi \nonumber\\
&+&\left. (\partial_z\Phi -\omega_H-\omega_h\cos\omega_{\rm p}t)\psi
\right\},
\label{eq:LLn}
\end{eqnarray}
and
\begin{eqnarray}
 \nabla^2\Phi&=&(1-2\psi\psi^*)
\left[ (\partial_x-\mathrm{i}\partial_y)\psi
 +(\partial_x+\mathrm{i}\partial_y)\psi^* \right] \nonumber\\
&-&2(\psi\partial_z\psi^*+\psi^*\partial_z\psi)
-\psi^{*2}(\partial_x+\mathrm{i}\partial_y)\psi
 -\psi^2(\partial_x-\mathrm{i}\partial_y)\psi^*.
\label{eq:poin}
\end{eqnarray}
Our interest lies in the region of static fields below the cusp point, 
where $\theta_{\vec{\scriptstyle k}}=\pi /2$. Then the expansion of $\psi$
is
\begin{eqnarray}
 \psi (\vec{r},t)&=&\sum_{\vec{\scriptstyle k}}
 a_{\vec{\scriptstyle k}}(t)
 \e^{\mathrm{i}(k_x x+k_y y)};\nonumber\\
 \psi^* (\vec{r},t)&=&\sum_{\vec{\scriptstyle k}}
 a^*_{-\vec{\scriptstyle k}}(t)
 \e^{\mathrm{i}(k_x x+k_y y)}.
\label{eq:expan_0} 
\end{eqnarray}
Substituting Eq.~(\ref{eq:expan_0}) into Eq.~(\ref{eq:poin}), we have
\begin{eqnarray}
 \nabla^2\Phi&=&\mathrm{i}\sum_{\vec{\scriptstyle k}}
 \e^{\mathrm{i}(k_x x+k_y y)}
(k_{-}a_{\vec{\scriptstyle k}}+k_{+}a^*_{-\vec{\scriptstyle k}})
\nonumber\\
&-&\mathrm{i}
\sum_{\vec{\scriptstyle k}_1\vec{\scriptstyle k}_2\vec{\scriptstyle k}_3}
\e^{\mathrm{i}[(k_{1x}+k_{2x}+k_{3x})x+(k_{1y}+k_{2y}+k_{3y})y]}
\nonumber\\
&& \quad \times \left[ (k_{3+}+2k_{1+}) a^*_{-\vec{\scriptstyle k}_1}
a^*_{-\vec{\scriptstyle k}_2}a_{\vec{\scriptstyle k}_3} +
(k_{3-}+2k_{2-})a_{\vec{\scriptstyle k}_1} a_{\vec{\scriptstyle k}_2}
a^*_{-\vec{\scriptstyle k}_3} \right] .
\label{eq:poin2}
\end{eqnarray}
It is convenient to restrict wave vectors for solving this equation. 
When $\vec{k}_1+\vec{k}_2+\vec{k}_3=\vec{k}$, the solution of
Eq.~(\ref{eq:poin2}) is  
\begin{eqnarray}
 \Phi&=&-\mathrm{i}\sum_{\vec{\scriptstyle k}}
 \frac{\e^{\mathrm{i}(k_x x+k_y y)}}{k^2}\left\{
(k_{-}a_{\vec{\scriptstyle k}}+k_{+}a^*_{-\vec{\scriptstyle k}})\right.
 \nonumber\\
&+&\sum_{\vec{\scriptstyle k}_1\vec{\scriptstyle k}_2
\vec{\scriptstyle k}_3} \left.
\left[ (k_{3+}+2k_{1+}) a^*_{-\vec{\scriptstyle k}_1}
a^*_{-\vec{\scriptstyle k}_2}a_{\vec{\scriptstyle k}_3} +
(k_{3-}+2k_{2-})a_{\vec{\scriptstyle k}_1} a_{\vec{\scriptstyle k}_2}
a^*_{-\vec{\scriptstyle k}_3} \right]
\right\} f_{\vec{\scriptstyle k}},
\label{eq:Phin}
\end{eqnarray}
where
\begin{equation}
 f_{\vec{\scriptstyle k}}=1-\frac{1-\e^{-k_\perp}}{k}.
\end{equation}
The derivation of these equations is similar to that of the linearized
equations. 

The nonlinear terms contributing to the resonance 
have spin waves whose wave vector is $\vec{k}$ or $-\vec{k}$. 
When the linear terms have $\vec{k}$, a possible combination of
$\vec{k}_1$, $\vec{k}_2$ and $\vec{k}_3$ is $(\vec{k},\vec{k}',-\vec{k}')$.
Here, however, we assume $\vec{k}_1,\vec{k}_2,\vec{k}_3=\pm\vec{k}$.
This assumption, which forbids the multi-mode couplings of spin
waves, incorporates the substantial nonlinear terms. 
Then, from Eqs.~(\ref{eq:LLn}), (\ref{eq:expan_0}) and (\ref{eq:Phin}),
we obtain
\begin{eqnarray}
 \partial_t a_{\vec{\scriptstyle k}}&+&
 \mathrm{i}(1-\mathrm{i}\lambda )\left\{
A_{\vec{\scriptstyle k}}a_{\vec{\scriptstyle k}}
+B_{\vec{\scriptstyle k}}a^*_{-\vec{\scriptstyle k}}
-\omega_h \cos(\omega_{\rm p}t)a_{\vec{\scriptstyle k}}
\right.\nonumber\\
&&+\left. 2(C_{\vec{\scriptstyle k}}+f_{\vec{\scriptstyle k}})
a_{\vec{\scriptstyle k}}a_{-\vec{\scriptstyle k}}
a^*_{-\vec{\scriptstyle k}}
+(-C_{\vec{\scriptstyle k}}+f_{\vec{\scriptstyle k}})
a_{\vec{\scriptstyle k}}a_{\vec{\scriptstyle k}}
a^*_{\vec{\scriptstyle k}}\right.\nonumber\\
&&+\left. {\textstyle \frac32}f_{\vec{\scriptstyle k}}
\e^{-2\mathrm{i}\varphi_{\vec{\scriptstyle k}}}a_{\vec{\scriptstyle k}}
a_{\vec{\scriptstyle k}}a_{-\vec{\scriptstyle k}}
+{\textstyle \frac12}f_{\vec{\scriptstyle k}}
\e^{2\mathrm{i}\varphi_{\vec{\scriptstyle k}}}
(2a^*_{-\vec{\scriptstyle k}}a^*_{\vec{\scriptstyle k}}
a_{\vec{\scriptstyle k}}+a^*_{-\vec{\scriptstyle k}}
a^*_{-\vec{\scriptstyle k}}a_{-\vec{\scriptstyle k}})
\right\}=0,
\label{eq:non_a}
\end{eqnarray}
where
\begin{eqnarray}
 A_{\vec{\scriptstyle k}}&=&1-l^2k^2-\omega_H-\frac12
f_{\vec{\scriptstyle k}},\nonumber\\
B_{\vec{\scriptstyle k}}&=&
-\frac12\e^{2\mathrm{i}\varphi_{\vec{\scriptstyle k}}}
f_{\vec{\scriptstyle k}}, \quad
C_{\vec{\scriptstyle k}}=-2l^2k^2.
\end{eqnarray}
As in the case of linear instability, we perform the
Holstein-Primakoff transformation~(\ref{eq:H-P}) and substitute
Eq.~(\ref{eq:bo}). When only terms contributing to the resonance are
left, the equations of motion for $b^o_{\vec{\scriptstyle k}}$ and
$b^{o*}_{-\vec{\scriptstyle k}}$ are
\begin{eqnarray}
 \partial_t b^o_{\vec{\scriptstyle k}}&-&\mathrm{i}\left(
\omega_{\vec{\scriptstyle k}}-{\textstyle \frac{\omega_{\rm p}}{2}}
+\mathrm{i}\eta_{\vec{\scriptstyle k}}\right)b^o_{\vec{\scriptstyle k}}
+\mathrm{i}\xi_{\vec{\scriptstyle k}}|b^o_{\vec{\scriptstyle k}}|^2
b^o_{\vec{\scriptstyle k}}+2\mathrm{i}\zeta_{\vec{\scriptstyle k}}
|b^o_{-\vec{\scriptstyle k}}|^2 b^o_{\vec{\scriptstyle k}}
-\rho_{\vec{\scriptstyle k}}b^{o*}_{-\vec{\scriptstyle k}}=0,
\nonumber\\
 \partial_t b^{o*}_{-\vec{\scriptstyle k}}&+&\mathrm{i}\left(
\omega_{\vec{\scriptstyle k}}-{\textstyle \frac{\omega_{\rm p}}{2}}
-\mathrm{i}\eta_{\vec{\scriptstyle k}}\right)
b^{o*}_{-\vec{\scriptstyle k}}
-\mathrm{i}\xi^*_{\vec{\scriptstyle k}}|b^{o}_{-\vec{\scriptstyle k}}|^2
b^{o*}_{-\vec{\scriptstyle k}}-2\mathrm{i}\zeta^*_{\vec{\scriptstyle k}}
|b^o_{\vec{\scriptstyle k}}|^2 b^{o*}_{-\vec{\scriptstyle k}}
-\rho^*_{\vec{\scriptstyle k}}b^o_{\vec{\scriptstyle k}}=0,
\label{eq:non_bo}
\end{eqnarray}
where
\begin{eqnarray}
 \xi_{\vec{\scriptstyle k}}&=&C_{\vec{\scriptstyle k}}\left\{
\frac{|B_{\vec{\scriptstyle k}}|^2(1+\lambda^2)}
{2\omega_{\vec{\scriptstyle k}}^2}-1
-\frac{\mathrm{i}\eta_{\vec{\scriptstyle k}}}
{\omega_{\vec{\scriptstyle k}}} \right\}\nonumber\\
&&+f_{\vec{\scriptstyle k}}\left\{ 1
+\frac{3|B_{\vec{\scriptstyle k}}|^2(1+\lambda^2)}
{2\omega_{\vec{\scriptstyle k}}^2}
+\frac{\mathrm{i}\eta_{\vec{\scriptstyle k}}}
{\omega_{\vec{\scriptstyle k}}}
-\frac{3|A_{\vec{\scriptstyle k}}|f_{\vec{\scriptstyle k}}\lambda^2}
{4\omega_{\vec{\scriptstyle k}}^2}
-\frac{3\mathrm{i}\lambda f_{\vec{\scriptstyle k}}}
{4\omega_{\vec{\scriptstyle k}}} \right\},\nonumber\\
 \zeta_{\vec{\scriptstyle k}}&=&C_{\vec{\scriptstyle k}}\left\{
\frac{|B_{\vec{\scriptstyle k}}|^2(1+\lambda^2)}
{2\omega_{\vec{\scriptstyle k}}^2}+1
+\frac{\mathrm{i}\eta_{\vec{\scriptstyle k}}}
{\omega_{\vec{\scriptstyle k}}} \right\}\nonumber\\
&&+f_{\vec{\scriptstyle k}}\left\{ 1
+\frac{3|B_{\vec{\scriptstyle k}}|^2(1+\lambda^2)}
{2\omega_{\vec{\scriptstyle k}}^2}
+\frac{\mathrm{i}\eta_{\vec{\scriptstyle k}}}
{\omega_{\vec{\scriptstyle k}}}
-\frac{3|A_{\vec{\scriptstyle k}}|f_{\vec{\scriptstyle k}}\lambda^2}
{4\omega_{\vec{\scriptstyle k}}^2}
-\frac{3\mathrm{i}\lambda f_{\vec{\scriptstyle k}}}
{4\omega_{\vec{\scriptstyle k}}} \right\}.
\end{eqnarray}
The detailed derivation is given in Appendix~\ref{sec:app_b}.
 
The time-independent solutions (fixed points) of Eq.~(\ref{eq:non_bo}),
$b^o_{\vec{\scriptstyle k}}=\overline{W}$ and
$b^{o*}_{-\vec{\scriptstyle k}}=\overline{W}^*$, are given by
\begin{equation}
 \left|\overline{W}\right|^2=\frac
{\eta_{\vec{\scriptstyle k}}\mathrm{Im}
(\xi_{\vec{\scriptstyle k}}+2\zeta_{\vec{\scriptstyle k}})\pm
\left[ \eta_{\vec{\scriptstyle k}}^2
\{\mathrm{Im}(\xi_{\vec{\scriptstyle k}}
 +2\zeta_{\vec{\scriptstyle k}})\}^2
+\left| \xi_{\vec{\scriptstyle k}}
 +2\zeta_{\vec{\scriptstyle k}}\right|^2
\left( \left|\rho_{\vec{\scriptstyle k}}\right|^2
- \eta_{\vec{\scriptstyle k}}^2 \right) \right]^{1/2}}
{\left| \xi_{\vec{\scriptstyle k}}
 +2\zeta_{\vec{\scriptstyle k}}\right|^2}.
\label{eq:W}
\end{equation}
Since the inside of the square root in Eq.~(\ref{eq:W}) should be
positive, 
\begin{equation}
 \left|\rho_{\vec{\scriptstyle k}}\right|^2 \ge \frac{[\mathrm{Re}
(\xi_{\vec{\scriptstyle k}}+2\zeta_{\vec{\scriptstyle k}})]^2}
{\left| \xi_{\vec{\scriptstyle k}}
 +2\zeta_{\vec{\scriptstyle k}}\right|^2}\eta_{\vec{\scriptstyle k}}^2.
\label{eq:insta_con}
\end{equation}
When 
$|\rho_{\vec{\scriptstyle k}}|^2>\eta_{\vec{\scriptstyle k}}^2$,
the double sign changes into the plus sign because 
$\left|\overline{W}\right|^2\ge 0$.
By using $b^o_{\vec{\scriptstyle k}}=\overline{W}+b_1$ and
$b^{o*}_{-\vec{\scriptstyle k}}=\overline{W}^*+b_2$ in
Eq.~(\ref{eq:non_bo}),  the linear equations of
$b_1$ and $b_2$  are written as
\begin{equation}
 \frac{\d}{\d t}\left(
\begin{array}{l}
 b_1\\ b_2
\end{array}
\right) = -\left[
\begin{array}{cc}
 \eta_{\vec{\scriptstyle k}}+2\mathrm{i}
(\xi_{\vec{\scriptstyle k}}+2\zeta_{\vec{\scriptstyle k}})
\left|\overline{W}\right|^2 & \mathrm{i}
(\xi_{\vec{\scriptstyle k}}+2\zeta_{\vec{\scriptstyle k}})
\overline{W}^2-\rho_{\vec{\scriptstyle k}}\\
-\mathrm{i}(\xi^*_{\vec{\scriptstyle k}}+2\zeta^*_{\vec{\scriptstyle k}})
\overline{W}^{*2}-\rho_{\vec{\scriptstyle k}}^* & \quad
\eta_{\vec{\scriptstyle k}}-2\mathrm{i}
(\xi^*_{\vec{\scriptstyle k}}+2\zeta^*_{\vec{\scriptstyle k}})
\left|\overline{W}\right|^2
\end{array}
\right] \left(
\begin{array}{l}
 b_1\\ b_2
\end{array}
\right) ,
\end{equation}
where we assumed $b_1^*=b_2$ and $b_2^*=b_1$.
The eigenvalues of the matrix are given as
\begin{eqnarray}
 \Lambda_{+}&=&-\eta_{\vec{\scriptstyle k}}+2\mathrm{Im}
(\xi_{\vec{\scriptstyle k}}+2\zeta_{\vec{\scriptstyle k}})
\left|\overline{W}\right|^2 +\left[ \eta_{\vec{\scriptstyle k}}^2
-4\{\mathrm{Re}
(\xi_{\vec{\scriptstyle k}}+2\zeta_{\vec{\scriptstyle k}}) \}^2
\left|\overline{W}\right|^4 \right]^{1/2}, \nonumber\\
 \Lambda_{-}&=&-\eta_{\vec{\scriptstyle k}}+2\mathrm{Im}
(\xi_{\vec{\scriptstyle k}}+2\zeta_{\vec{\scriptstyle k}})
\left|\overline{W}\right|^2 -\left[ \eta_{\vec{\scriptstyle k}}^2
-4\{\mathrm{Re}
(\xi_{\vec{\scriptstyle k}}+2\zeta_{\vec{\scriptstyle k}}) \}^2
\left|\overline{W}\right|^4 \right]^{1/2}.
\end{eqnarray}
Since $\Lambda_{+}>\Lambda_{-}$, the instability condition is
$\Lambda_{+}>0$. However, this condition does not lead to any nontrivial
condition. Then the actual instability condition is
Eq.~(\ref{eq:insta_con}). The instability threshold now becomes
\begin{equation}
 \omega_h^{\rm crit}=\frac{\omega_{\rm p}\eta_{\vec{\scriptstyle k}}}
{\left| B_{\vec{\scriptstyle k}}\right|\sqrt{1+\lambda^2}}
\frac{\left| \mathrm{Re}
(\xi_{\vec{\scriptstyle k}}+2\zeta_{\vec{\scriptstyle k}})\right|}
{\left|\xi_{\vec{\scriptstyle k}}+2\zeta_{\vec{\scriptstyle k}}\right|}.
\label{wq:nomegah}
\end{equation}
However, this equation cannot explain the experimental butterfly curve
of the instability threshold, either. The factor $|\mathrm{Re}
(\xi_{\vec{\scriptstyle k}}+2\zeta_{\vec{\scriptstyle k}})|/
|\xi_{\vec{\scriptstyle k}}+2\zeta_{\vec{\scriptstyle k}}|$ is almost
constant for the region where our calculation is performed, and we
cannot overcome the problem of divergence 
$\omega_h^{\rm crit}\to\infty$ as $k\to 0$. 

\section{Discussion}

We have tried to derive the spin-wave instability threshold for parallel
pumping from the LL equation. Neither the linear nor 
nonlinear analysis 
has proved to explain the feature of experimental butterfly
curves of the threshold. By contrast, the Suhl instability for
perpendicular pumping can be well explained from the LL
equation~\cite{Suhl}. The successful explanation of Suhl instability is
based on the fact that the couplings between the uniform mode and other
spin-wave 
modes are dominant for perpendicular pumping. In parallel pumping,
however, the uniform mode is not excited. That is the reason why the
Suhl instability cannot be applied to the present case.

In order to explain the instability threshold for parallel pumping, it
is necessary to deal with the spin-wave relaxation in a proper way. 
Conventionally, the spin-wave line width $\Delta H_k$, which is
calculated from the microscopic theory, is used to describe the
spin-wave relaxation. In Ref.~\cite{Patton}, the $k$-dependence of 
$\Delta H_k$ plays a vital role in obtaining the butterfly curve. Such a
$k$-dependence of $\Delta H_k$ should correspond to the multi-mode 
spin-wave coupling, which is not considered in the present treatment of
the LL 
equation. However, the equation of motion with multi-mode couplings is
too complex to be solved analytically. 
That may be the reason why the instability threshold for
parallel pumping has never been derived in the context of nonlinear
dynamics by using only the LL equation.

\section*{Acknowledgments}

The authors thank to M.~Mino of Okayama university for useful
discussion. One of the authors (K. K.) is supported by JSPS Research
Fellowships for Young Scientists.

\appendix

\section{\label{sec:app} Demagnetization field}

Substituting Eqs.~(\ref{eq:expan_a}) and (\ref{eq:expan_b}) into the
Poisson equation, we have
\begin{eqnarray}
 \nabla^2\Phi &=&\mathrm{i}\sum_{\vec{\scriptstyle k}}
 \e^{\mathrm{i}(k_x x+k_y y)}
 (k_{-}a_{\vec{\scriptstyle k}}+k_{+}a^*_{-\vec{\scriptstyle k}})
 \cos k_z z\quad\mbox{for even modes};\nonumber\\
 \nabla^2\Phi &=&\mathrm{i}\sum_{\vec{\scriptstyle k}}
 \e^{\mathrm{i}(k_x x+k_y y)}
 (k_{-}a_{\vec{\scriptstyle k}}-k_{+}a^*_{-\vec{\scriptstyle k}})
 \sin k_z z\quad\mbox{for odd modes}.
\label{eq:A-poi}
\end{eqnarray}
Generally, for the Poisson equation given as
\begin{equation}
 \nabla^2\phi(\vec{r})=-4\pi\rho (\vec{r}), 
\end{equation}
the solution is
\begin{equation}
 \phi(\vec{r})=\int\frac{\rho (\vec{r}')\d
  \vec{V}'}{|\vec{r}'-\vec{r}|}. 
\label{eq:a_phi}
\end{equation}
By using Eq.~(\ref{eq:a_phi}), the solutions of Eq.{~(\ref{eq:A-poi})
are written as
\begin{eqnarray}
  \Phi(\vec{r})=-\frac{\mathrm{i}}{4\pi}\sum_{\vec{\scriptstyle k}}
(k_{-}a_{\vec{\scriptstyle k}}+k_{+}a^*_{-\vec{\scriptstyle k}})
&\int&\frac{\d \vec{V}'}{|\vec{r}'-\vec{r}|}
\e^{\mathrm{i}(k_x x'+k_y y')}\cos k_z z' \nonumber\\
&&\quad\quad\mbox{for even mode};\nonumber\\
  \Phi(\vec{r})=-\frac{\mathrm{i}}{4\pi}\sum_{\vec{\scriptstyle k}}
(k_{-}a_{\vec{\scriptstyle k}}-k_{+}a^*_{-\vec{\scriptstyle k}})
&\int&\frac{\d \vec{V}'}{|\vec{r}'-\vec{r}|}
\e^{\mathrm{i}(k_x x'+k_y y')}\sin k_z z' \nonumber\\
&&\quad\quad\mbox{for odd mode}.
\label{eq:A4}
\end{eqnarray}
After integration, Eq.~(\ref{eq:A4}) becomes
\begin{eqnarray}
  \Phi(\vec{r})=-\mathrm{i}\sum_{\vec{\scriptstyle k}}
\frac{\e^{\mathrm{i}(k_x x+k_y y)}}{k^2}
(k_{-}a_{\vec{\scriptstyle k}}+k_{+}a^*_{-\vec{\scriptstyle k}})&&
\left[ \cos k_z z-(-1)^m\e^{-k_\perp /2}\cosh k_\perp z \right]
\nonumber\\
&&\quad\quad\mbox{for even mode};\nonumber\\
  \Phi(\vec{r})=-\mathrm{i}\sum_{\vec{\scriptstyle k}}
\frac{\e^{\mathrm{i}(k_x x+k_y y)}}{k^2}
(k_{-}a_{\vec{\scriptstyle k}}-k_{+}a^*_{-\vec{\scriptstyle k}})&&
\left[ \sin k_z z-(-1)^m\e^{-k_\perp /2}\sinh k_\perp z \right]
\nonumber\\
&&\quad\quad\mbox{for odd mode}.
\end{eqnarray}
Now let us introduce the projection $F(z)$ of an arbitrary
function $f(z)$ onto $\cos k_z z$: 
\begin{equation}
 F(z)=\int_{-1/2}^{1/2}\d z
 f(z)\cos k_z z
  \left/ \int_{-1/2}^{1/2}\d z \cos^2 k_z z\right. ,
\end{equation}
for  $-1/2 < z < 1/2$. The projection onto $\sin k_z z$ is also defined in
the same way.
Applying the above projection, we obtain 
\begin{eqnarray}
 \Phi (\vec{r})&=&-\mathrm{i}\sum_{\vec{\scriptstyle k}}
 \frac{\e^{\mathrm{i}(k_x x+k_y y)}}{k^2}
(k_{-}a_{\vec{\scriptstyle k}} +k_{+}a^*_{-\vec{\scriptstyle k}})
f_{\vec{\scriptstyle k}}\cos k_z z
\quad \mbox{for even modes;} \nonumber\\
 \Phi (\vec{r})&=&-\mathrm{i}\sum_{\vec{\scriptstyle k}}
 \frac{\mathrm{e}^{\mathrm{i}(k_x x+k_y y)}}{k^2}
(k_{-}a_{\vec{\scriptstyle k}} -k_{+}a^*_{-\vec{\scriptstyle k}})
f_{\vec{\scriptstyle k}}\sin k_z z
\quad \mbox{for odd modes.}
\end{eqnarray}
Here,
\begin{eqnarray}
 f_{\vec{\scriptstyle k}}&=&1-\left(
1-\e^{-k_{\perp}} \right) \frac{k_{\perp}}{k^2}
\quad \mbox{when $k_z=0$;} \nonumber\\
 f_{\vec{\scriptstyle k}}&=&1-2\left(
1-\e^{-k_{\perp}} \right) \frac{k_{\perp}}{k^2}
\quad \mbox{when $k_z\ne 0$.}
\end{eqnarray}

\section{\label{sec:app_b} Nonlinear equation of motion for
 $b^o_{\vec{\scriptstyle k}}$} 

First of all, let the terms of Eq.~(\ref{eq:non_a}) be represented by 
$P_{1,\vec{\scriptstyle k}}$, $P_{2,\vec{\scriptstyle k}}$, 
$P_{3,\vec{\scriptstyle k}}$, $P_{4,\vec{\scriptstyle k}}$ and
$P_{5,\vec{\scriptstyle k}}$:
\begin{eqnarray}
P_{1,\vec{\scriptstyle k}}&\equiv& \partial_t a_{\vec{\scriptstyle k}}
+\mathrm{i}(1-\mathrm{i}\lambda )\left(
A_{\vec{\scriptstyle k}}a_{\vec{\scriptstyle k}}
+B_{\vec{\scriptstyle k}}a^*_{-\vec{\scriptstyle k}}
-\omega_h \cos(\omega_{\rm p}t) a_{\vec{\scriptstyle k}}
\right)\nonumber\\
P_{2,\vec{\scriptstyle k}}&\equiv&2\mathrm{i}(1-\mathrm{i}\lambda )
(C_{\vec{\scriptstyle k}}+f_{\vec{\scriptstyle k}})
a_{\vec{\scriptstyle k}}a_{-\vec{\scriptstyle k}}
a^*_{-\vec{\scriptstyle k}}\nonumber\\
P_{3,\vec{\scriptstyle k}}&\equiv&\mathrm{i}(1-\mathrm{i}\lambda )
(-C_{\vec{\scriptstyle k}}+f_{\vec{\scriptstyle k}})
a_{\vec{\scriptstyle k}}a_{\vec{\scriptstyle k}}
a^*_{\vec{\scriptstyle k}}\nonumber\\
P_{4,\vec{\scriptstyle k}}&\equiv&
{\textstyle \frac32}\mathrm{i}(1-\mathrm{i}\lambda )
f_{\vec{\scriptstyle k}}
\e^{-2\mathrm{i}\varphi_{\vec{\scriptstyle k}}}a_{\vec{\scriptstyle k}}
a_{\vec{\scriptstyle k}}a_{-\vec{\scriptstyle k}}\nonumber\\
P_{5,\vec{\scriptstyle k}}&\equiv&
{\textstyle \frac12}\mathrm{i}(1-\mathrm{i}\lambda )
f_{\vec{\scriptstyle k}}
\e^{2\mathrm{i}\varphi_{\vec{\scriptstyle k}}}
(2a^*_{-\vec{\scriptstyle k}}a^*_{\vec{\scriptstyle k}}
a_{\vec{\scriptstyle k}}+a^*_{-\vec{\scriptstyle k}}
a^*_{-\vec{\scriptstyle k}}a_{-\vec{\scriptstyle k}}).
\end{eqnarray}  
Next, we introduce a new quantity 
$R_{i,\vec{\scriptstyle k}}\equiv\nu_{\vec{\scriptstyle k}}
P_{i,\vec{\scriptstyle k}}+\mu_{\vec{\scriptstyle k}} 
P^*_{i,-\vec{\scriptstyle k}}$ for $i=1,\ldots,5$.
After the Holstein-Primakoff transformation~(\ref{eq:H-P}), we use
Eq.~(\ref{eq:bo}) and remove the terms which do not contribute to the
resonance. Then, we have
\begin{eqnarray}
 R_{1,\vec{\scriptstyle k}}&=&\partial_t b^o_{\vec{\scriptstyle k}}
-\mathrm{i}\left(
\omega_{\vec{\scriptstyle k}}-{\textstyle \frac{\omega_{\rm p}}{2}}
+\mathrm{i}\eta_{\vec{\scriptstyle k}} \right) b^o_{\vec{\scriptstyle k}} 
-\rho_{\vec{\scriptstyle k}}b^{o*}_{-\vec{\scriptstyle k}}\nonumber\\
R_{2,\vec{\scriptstyle k}}&=&2\mathrm{i}
(C_{\vec{\scriptstyle k}}+f_{\vec{\scriptstyle k}})
\left\{\left( \frac{A_{\vec{\scriptstyle k}}^2}
{\omega_{\vec{\scriptstyle k}}^2}+
\frac{\mathrm{i}\eta_{\vec{\scriptstyle k}}}
{\omega_{\vec{\scriptstyle k}}}\right) 
|b^o_{-\vec{\scriptstyle k}}|^2 +
\frac{\left|B_{\vec{\scriptstyle k}}\right|^2(1+\lambda^2)}
{2\omega_{\vec{\scriptstyle k}}^2}
|b^o_{\vec{\scriptstyle k}}|^2 \right\}
b^o_{\vec{\scriptstyle k}}\nonumber\\
R_{3,\vec{\scriptstyle k}}&=&\mathrm{i}
(-C_{\vec{\scriptstyle k}}+f_{\vec{\scriptstyle k}})
\left\{ \left( 1+
\frac{\left|B_{\vec{\scriptstyle k}}\right|^2(1+\lambda^2)}
{2\omega_{\vec{\scriptstyle k}}^2}
+\frac{\mathrm{i}\eta_{\vec{\scriptstyle k}}}
{\omega_{\vec{\scriptstyle k}}}\right)
|b^o_{\vec{\scriptstyle k}}|^2
+\frac{\left|B_{\vec{\scriptstyle k}}\right|^2(1+\lambda^2)}
{\omega_{\vec{\scriptstyle k}}^2}
|b^o_{-\vec{\scriptstyle k}}|^2\right\}
b^o_{\vec{\scriptstyle k}}\nonumber\\
R_{4,\vec{\scriptstyle k}}&=&
-\frac{3\mathrm{i}f_{\vec{\scriptstyle k}}^2}
{8\omega_{\vec{\scriptstyle k}}^2}
\left\{ -(1-\lambda^2)\frac{\left|A_{\vec{\scriptstyle k}}\right|}
{\omega_{\vec{\scriptstyle k}}}+2\mathrm{i}\lambda\right\}
\left( 2|b^o_{-\vec{\scriptstyle k}}|^2
+|b^o_{\vec{\scriptstyle k}}|^2\right)
b^o_{\vec{\scriptstyle k}}\nonumber\\
R_{5,\vec{\scriptstyle k}}&=&
-\frac{3\mathrm{i}\left|A_{\vec{\scriptstyle k}}\right| 
f_{\vec{\scriptstyle k}}(1+\lambda^2)}
{8\omega_{\vec{\scriptstyle k}}^2}
\left( 2|b^o_{-\vec{\scriptstyle k}}|^2
+|b^o_{\vec{\scriptstyle k}}|^2\right)
b^o_{\vec{\scriptstyle k}},
\label{eq:Rs}
\end{eqnarray}
where
\begin{equation}
 \rho_{\vec{\scriptstyle k}}=\mathrm{i}(1-\mathrm{i}\lambda )\omega_h
\frac{B_{\vec{\scriptstyle k}}}{2\omega_{\vec{\scriptstyle k}}}.
\end{equation}
The terms in Eq.~(\ref{eq:Rs}) satisfy the following:
\begin{equation}
 R_{1,\vec{\scriptstyle k}}+R_{2,\vec{\scriptstyle k}}+
R_{3,\vec{\scriptstyle k}}+R_{4,\vec{\scriptstyle k}}+
R_{5,\vec{\scriptstyle k}}=0.
\label{eq:BR}
\end{equation}
Equation~(\ref{eq:non_bo}) is obtained from Eqs.~(\ref{eq:Rs}) and
(\ref{eq:BR}).

\end{document}